\documentclass[12pt,preprint,eqsecnum,nofootinbib,amsmath,amssymb,floatfix]{revtex4}
\usepackage{graphicx}
\usepackage{dcolumn}
\usepackage{bm}
\usepackage{epsfig,psfrag,amsmath,amssymb,float}

\begin{document}
                                                                                
                                                                                
\def\leftrh{James B.~Hartle}
\def\rightrh{Quasiclassical Domains In A Quantum Universe}
                                                                                
\def\frac#1#2{{\textstyle #1 \over \textstyle #2}}
\def\half{{\textstyle {1 \over 2}}}
\def\viz{{\it viz.~}}
\def\eg{{\it e.g.~}}
\def\ie{{\it i.e.,\ }}
\def\gtwid{\mathrel{\raise.3ex\hbox{$>$\kern-.75em\lower1ex\hbox{$\sim$}}}}
\def\ltwid{\mathrel{\raise.3ex\hbox{$<$\kern-.75em\lower1ex\hbox{$\sim$}}}}
\def\ed{{\it ed.~by\ }}
                                                                                
\title{THE SPACETIME APPROACH TO QUANTUM MECHANICS\footnote{
This essay is dedicated to Louis Witten on the occasion of his
retirement from the University of Cinncinati.  It appeared in the
proceedings of the symposium held in his honor on April 4---5, 1992,
 {\it Topics on Quantum Gravity and Beyond}, ed. by F. Mansouri and J.J. Scanio,  in the Proceedings of the
{\sl International Symposium on Quantum Physics and the Universe},
Wasada University, Tokyo, Japan, August 23--27, 1992, and in 
Vistas in Astronomy {\bf 37}, 569, (1993)}}


\author{James B.~Hartle}
\email{hartle@physics.ucsb.edu}
\affiliation{Department of Physics,
University of California\\
 Santa Barbara, CA 93106-9530 USA}

\begin{abstract}

\setlength{\baselineskip}{.2in}

Feynman's sum-over-histories formulation of quantum mechanics is
reviewed as an independent statement of quantum theory in spacetime
form.  It is different from the usual Schr\"odinger-Heisenberg
formulation that utilizes states on  spacelike surfaces because it assigns
probabilities to different sets of alternatives. In a sum-over-histories
formulation, alternatives at definite
moments of time are more restricted than
in usual quantum mechanics because they refer only to the coordinates
in terms of which the histories are defined.
However, in the context of the
quantum mechanics of closed systems, sum-over-histories quantum
mechanics can be generalized to deal with spacetime alternatives that
are not ``at definite moments of time''. An example in field theory
is the set of
alternative ranges of values of a field averaged over a spacetime
region.  An example in particle mechanics is the set of
the alternatives defined by
whether a particle never crosses a fixed spacetime region or crosses it
at least once.  The general notion of a set of  spacetime alternatives is a
partition (coarse-graining) of the histories into
 an exhaustive set of exclusive classes.  With this generalization the
sum-over-histories formulation can be said to be in fully spacetime form
with dynamics represented by path integrals over spacetime histories and
alternatives defined  as spacetime partitions of these histories.
When restricted to alternatives at definite moments of times this
generalization is equivalent to Schr\"odinger-Heisenberg quantum
mechanics.  However, the quantum mechanics of more general spacetime
alternatives does not have an equivalent Schr\"odinger-Heisenberg
formulation.
We suggest
that, in the quantum theory of gravity, the general notion of
``observable'' is supplied by diffeomorphism invariant partitions of
spacetime metrics and matter field configurations. By generalizing the
usual alternatives so as to put quantum theory in fully
spacetime form we may be led to a covariant generalized quantum
mechanics of spacetime free from the problem of time.

\end{abstract}

\maketitle
                                                                                
\setlength{\baselineskip}{.3in}

\setcounter{footnote}{0}
\section{Introduction}
\label{sec:I}

In 1948 Feynman \cite{Fey48}, building on the work of
Dirac \cite{Dir33}, introduced his
sum-over-histories formulation of quantum mechanics and with it the path
integral that has proved a powerful tool in many branches of
physics.  It is possible to see the sum-over-histories as merely a
technical tool --- a useful device for computing certain amplitudes
within the usual Schr\"odinger--Heisenberg formulation of quantum
mechanics in terms of states on spacelike surfaces.
That is not, however, how I think Feynman saw it.
Rather, the sum-over-histories formulation of quantum mechanics can be
regarded as an independent formulation of quantum
theory.  Its equivalence with the usual formulations is not
automatic but rather a question
whose answer may be different in different theories and for different
physical systems.
In this talk I shall review the current
status of the sum-over-histories formulation as an independent statement
of quantum theory.  I shall argue that, viewed most fundamentally, it is
{\it different} from the Schr\"odinger--Heisenberg formulations because
the totality of alternatives to which
it potentially assigns probabilities is different from that of
Schr\"odinger-Heisenberg quantum mechanics.
Sum-over-histories
  alternatives at definite moments of time are more restricted
than in usual quantum mechanics because
they refer only to the coordinates of the configuration space
in which the paths are defined. By contrast, at a moment of time,
Schr\"odinger-Heisenberg quantum mechanics has all the alternatives
provided by transformation theory.
However, even in non-relativistic quantum mechanics, the
sum-over-histories formulation allows a generalization of the
alternatives at definite moments of time
to genuine spacetime alternatives that are not considered in usual
Schr\"odinger-Heisenberg formulations.  This generalization allows a more
realistic description of everyday experiments.  But, more
importantly, it may be central for the construction of a quantum theory
of gravity in which there is no well-defined notion of time.  There,
alternatives ``at a moment of time'' may be difficult to find, while
spacetime alternatives may be the natural ``observables'' for which the
theory makes predictions.

\section{The Spacetime Approach to Quantum Mechanics}

In the Schr\"odinger--Heisenberg quantum mechanics of particles or
fields moving in a fixed background spacetime, the quantum dynamics of
measured subsystems is
formulated in terms of state vectors defined on spacelike surfaces that
evolve unitarily in between measurements and are reduced at
measurements. Unitary evolution is represented by
\begin{equation}
\bigl | \psi(t)\bigr\rangle = e^{-iHt/\hbar}\bigr | \psi
\bigr\rangle
\label{twoone}
\end{equation}
where $H$ is the Hamiltonian and $|\psi\rangle$ is the state at $t=0$.
 From \eqref{twoone} we could calculate
the transition amplitude between coordinates
$q^\prime$ at time $t^\prime$ to coordinates $q^{\prime\prime}$ at
time $t^{\prime\prime}$, that is an equivalent summary of unitary
evolution, \viz
\begin{equation}
\bigl\langle q^{\prime\prime}t^{\prime\prime}\bigl | q^\prime
t^\prime\bigr\rangle = \bigl\langle q^{\prime\prime} \bigl |
e^{-iH(t^{\prime\prime} - t^\prime)/\hbar}\bigr | q^\prime
\bigr\rangle\, .
\label{twotwo}
\end{equation}
(Coordinate indices, which may refer to either particles or fields,
are often omitted to keep the notation compact).
In a sum-over-histories formulation of quantum mechanics such amplitudes,
 are specified {\it directly} as path integrals
\begin{equation}
\bigl\langle q^{\prime\prime} t^{\prime\prime}\bigr  | q^\prime
t^\prime\bigr\rangle = \int\nolimits_{[q^\prime, q^{\prime\prime}]}\delta q
\ \exp (i S[q(\tau)]/\hbar)\, . 
\label{twothree}
\end{equation}
Here, $S[q(\tau)]$ is the action functional corresponding to the
Hamiltonian $H$ and the sum is over paths $q(t)$ that start at
$q^\prime$ at time $t^\prime$, end at $q^{\prime\prime}$ at time
$t^{\prime\prime}$, and are single-valued functions of time.  In an
abbreviated notation, we may write
\begin{equation}
\int \delta q^\prime\ \exp\bigr(iS[q(\tau)]/\hbar\bigr) \bigr |\psi\bigr\rangle
\label{twofour}
\end{equation}
for the state vector $|\psi(t)\rangle$ that is evolved by the propagator
\eqref{twothree}. More explicitly, \eqref{twofour} stands for the state vector
$|\psi(t)\rangle$ whose representative wave function
$\psi(q,t) = \langle q|\psi(t)\rangle$ is
\begin{equation}
\psi (q,t) = \int dq^\prime \left(\int\nolimits_{[q^\prime, q]} \delta q
\ \exp\bigl(iS[q(\tau)]/\hbar\bigr)\right) \ \psi(q^\prime, t^\prime)
\, .
\label{twofive}
\end{equation}
In this way, quantum dynamics corresponding to unitary evolution is cast
into manifestly spacetime form involving spacetime histories directly.
This is an important advantage in dealing with spacetime
symmetries such as Lorentz invariance.

Unitary evolution, however, is not the only law by which the state
vector evolves in quantum mechanics.  In the usual discussion,
at an ideal measurement that
disturbs the measured system as little as possible, the state vector is
instantaneously ``reduced'' according to the ``second law of evolution''
\begin{equation}
\big | \psi (t) \bigr\rangle \longrightarrow \frac{P_\alpha\big|
\psi(t)\bigr\rangle}{\big\Arrowvert P_\alpha\big|
\psi(t)\bigr\rangle\big\Arrowvert}
\label{twosix}
\end{equation}
where $P_\alpha$ is the projection operator on the subspace
 corresponding to the outcome
of the measurement and $\parallel\cdot\parallel$ denotes the norm of a
vector in Hilbert space.  This ``second law of evolution'' may not be needed
to calculate the transition probabilities in scattering experiments but
it is essential for calculating the probabilities of the {\it sequences} of
observations that define the histories of everyday life such as the
orbit of the earth around the sun. It is every bit as
essential for the prediction of realistic probabilities as
is unitary evolution.

 Using the two laws of evolution,
the joint probability for a
sequence of measured outcomes $\alpha_1, \cdots, \alpha_n$ at times $t_1,
\cdots, t_n$ may be calculated.  It finds its most compact expression in
the Heisenberg picture:
\begin{equation}
\Big\Arrowvert P^n_{\alpha_n} (t_n) P^{n-1}_{\alpha_{n-1}} (t_{n-1})
\cdots P^1_{\alpha_1} (t_1) \bigr|\psi\bigr\rangle \Big\Arrowvert^2
\, .\label{twoseven}
\end{equation}
Here, $\{P^k_{\alpha_k} (t_k)\}$ is an exhaustive set of orthogonal
Heisenberg picture projections representing the alternatives $\alpha_k$
in the set $k$ at time $t_k$.  For example, the set of alternatives
might be an exhaustive
 set of alternative regions $\Delta^k_{\alpha_k}$ of the
coordinates $q$ at time $t_k$.

The second law of evolution can be expressed simply in
sum-over-histories form
if attention is restricted to alternatives defined by
sequences of configuration space regions $\{\Delta^1_{\alpha_1}\},
\{\Delta^2_{\alpha_2}\}, \cdots$ at times $t_1, \cdots, t_n$ 
\cite{Cav86,  Sta86}. The
joint probability that the system passes through the particular
sequence of regions
$\alpha = (\alpha_1, \cdots, \alpha_n)$ is
\begin{equation}
\Big\Arrowvert\int\nolimits_{c_\alpha} \delta q
\ \exp\bigl(iS[q(\tau)]/\hbar\bigr)\bigr|\psi\bigr\rangle \Big\Arrowvert^2
\label{twoeight}
\end{equation}
where the path integral is over the class of
 paths $c_\alpha$  that pass through the region
$\Delta^1_{\alpha_1}$ at time $t_1$, $\Delta^2_{\alpha_2}$ at time
$t_2$, etc.  Eq.~\eqref{twoeight} is a unified expression for the two
laws of evolution in quantum mechanics.

Eq.~\eqref{twoeight} is equivalent to \eqref{twoseven} for alternatives
defined by exhaustive sets of exclusive regions of
configuration space at definite
moments of time.   In this case the Schr\"odinger--Heisenberg
formulation of quantum mechanics and the sum-over-histories formulation
coincide.  But they are not {\it fully} equivalent because the effect of
projections onto ranges of momentum, for example, cannot be represented
as restrictions on a configuration space
 path integral as in \eqref{twoeight}.  Probabilities
for momentum measurements {\it can} be predicted using path integrals,
 but only
approximately, by modeling a time of flight determination of velocity in
configuration space terms \cite{FH65}.
The sum-over-histories formulation of
quantum mechanics therefore deals directly and exactly with a more
restricted class of alternatives at sequences of moments of time
than is available from the
transformation theory of the Schr\"odinger--Heisenberg formulation.

\section{Spacetime Alternatives}

As we have presented it so far, the sum-over-histories formulation of
quantum mechanics is not in fully spacetime form.  The use of path
integrals over spacetime histories has put the  dynamics
corresponding to unitary evolution into spacetime form. But the
alternatives to which the theory
assigns probabilities are not general spacetime
alternatives.  They have been restricted to sequences of alternative
regions
of configuration space at definite moments of time.  More
general spacetime alternatives are easy to imagine.  We mention
two: Consider the quantum mechanics of a particle in which the histories
are paths in spacetime.  Fix a spacetime region $R$ with extent both in
space and time (Figure 1).  A given particle path may never cross $R$
or, alternatively, it may cross $R$ sometime, perhaps more than once.
These are an exhaustive set of spacetime alternatives for the particle
that are not ``at a moment of time''.  A second example is provided by
field averages over spacetime regions with extent both in space and time
such as were considered by Bohr and Rosenfeld \cite{BR33} in their
discussion of the measurability of the electromagnetic field.  An
exhaustive set of ranges of such average values of a field is an example
of a set of spacetime alternatives in field theory. Such spacetime
alternatives are not directly assigned probabilities in
Schr\"odinger-Heisenberg quantum mechanics because they are not ``at a
moment in time''.

\begin{figure}[t]
\begin{center}
\includegraphics[width=4in]{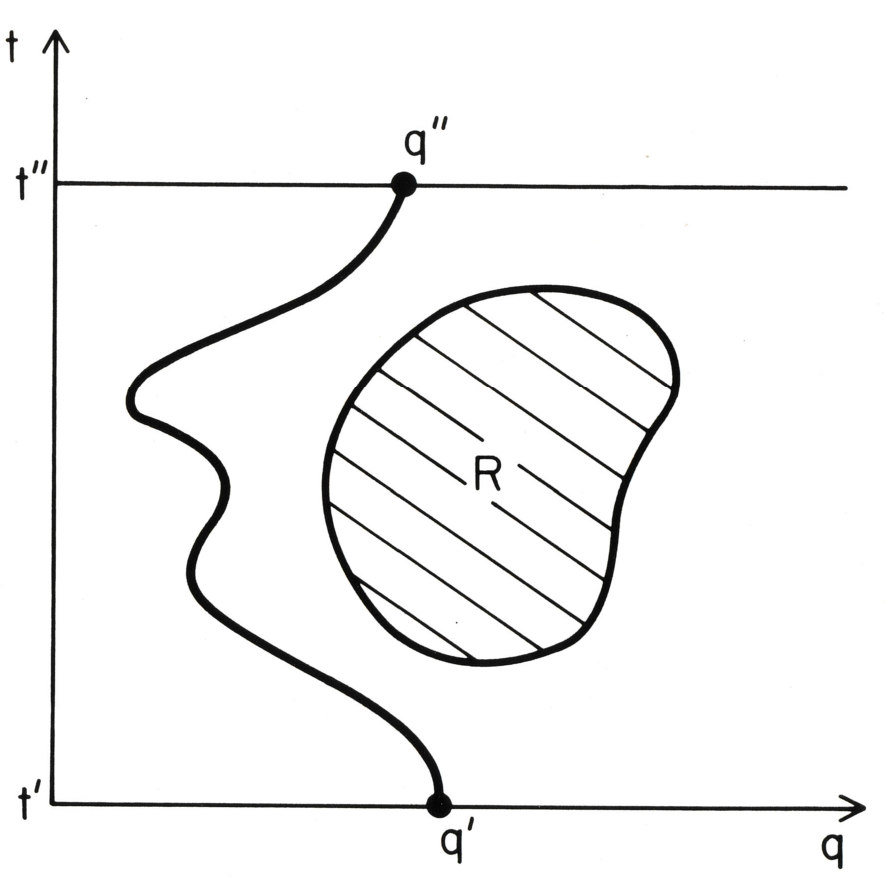}
\caption{Spacetime Alternatives defined by a
spacetime region.  The figure shows a spacetime region $R$ with extent
in both space and time.  The paths of a non-relativistic particle
between $q^\prime$ at time $t^\prime$ and $q^{\prime\prime}$ at time
$t^{\prime\prime}$ may be divided into two classes: First, the class of
paths that never cross the region $R$ (illustrated).  Second, the class
of paths that cross $R$ sometime.  These two classes constitute an
exhaustive set of alternatives for the particle that are not ``at a
moment of time''.}
\end{center}
\end{figure}

In his 1948 paper, Feynmann discussed alternatives defined with respect
to spacetime regions such as we described above.  In particular, he
offered a sum-over-histories definition of the probability that ``if an
ideal measurement is performed to determine whether a particle has a
path lying in a region of spacetime ...  the result will be
affirmative''.  However, that discussion, as well as more recent ones
\cite{Men79, Cav86, Schm87, Har88, Sor89,  YT91a}
were incomplete because they did not specify clearly what such an ideal
measurement consisted of or what was to replace the ``second law of
evolution'', \eqref{twosix}, following its completion.  ``I have not been able
to find a precise definition'' Feynman said \cite{Fey48}.
Only recently has it become
clear how to generalize sum-over-histories quantum mechanics to predict
probabilities for such spacetime alternatives within the quantum
mechanics of closed systems in which the notion of ``measurement'' does
not play a fundamental role \cite{Har91, Har91b, YT91b, YT92}.
  I shall discuss this generalization, but
first we must review a bit of the quantum mechanics of closed systems.

\section{The Quantum Mechanics of Closed Systems}

The most general objective of quantum theory is to predict the
probabilities of individual histories in an exhaustive set
of alternative, coarse-grained histories of a closed system.
 A characteristic feature of a
quantum-mechanical theory is that not every set of histories that may be
described can be assigned probabilities because of quantum-mechanical
interference between
the individual histories in the set.
Nowhere is this more clearly illustrated than in the
two-slit experiment (Figure 2).  In the
usual discussion, if we have not
measured which slit the electron went through on its way to being
detected at the screen, then we are not permitted to assign
probabilities to the alternative histories in which it passed through
the upper or lower slit.  It would be inconsistent to do so since the
correct probability sum rules would not be satisfied.  Because of
interference, the probability to arrive at a point $y$ on the screen is
not the sum of the probabilities to arrive at $y$ going through the
upper and the lower slit:

\begin{figure}[t]
\begin{center}
\includegraphics[width=4in]{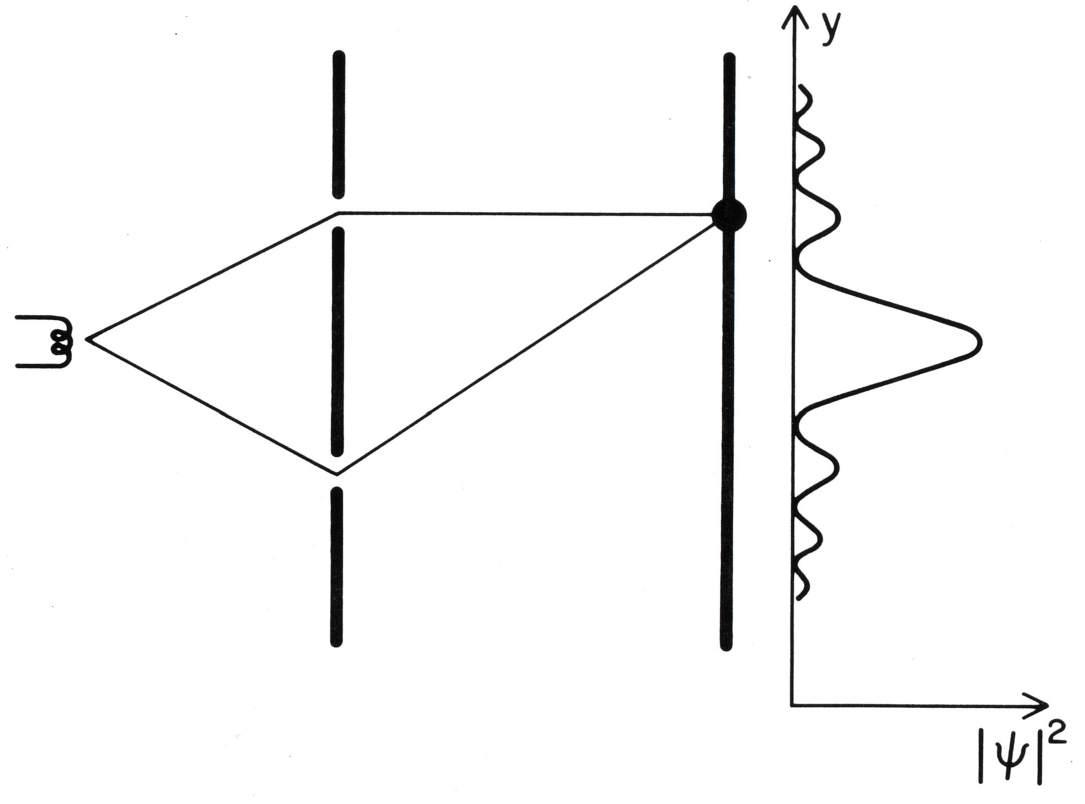}
\caption{The two-slit experiment.  An
electron gun at left emits an electron traveling towards a screen with
two slits, its progress in space recapitulating its evolution in time.  When
precise detections are made of an ensemble of such electrons at the
screen it
is not possible, because of interference, to assign a probability to the
alternatives of whether an individual electron went through the upper
slit or
the lower slit.  However, if the electron interacts with apparatus that
measures which slit it passed through, then these alternatives decohere
and
probabilities can be assigned.}
\end{center}
\end{figure}

\begin{equation}
\bigl|\psi_L (y) + \psi_U (y) \bigr|^2 \not= \bigl| \psi_L (y) \bigr|^2
+ \bigl|\psi_U (y) \bigr|^2\, . 
\label{fourone}
\end{equation}
In quantum mechanics a rule is needed to determine which sets of
histories can be assigned probabilities and then what those
probabilities are.

In the ``Copenhagen''
 quantum mechanics of measured subsystems probabilities can be
assigned to alternative histories that have been {\it measured}.  In the two
slit experiment, for example, if we {\it have} measured which slit the
electron went through, then interference is destroyed, the sum rule
obeyed and we {\it can} consistently assign probabilities to the
alternative histories in which the electron passed through the upper or
lower slit.

In the quantum mechanics of closed systems, containing both observer and
observed, measuring apparatus and measured subsystem, the above rule is
but a special case of a more general one of much wider
applicability \cite{Gri84, Omn88, GH90, YT91b}.
Probabilities can be assigned to the individual members of a set of
alternative coarse-grained histories of a closed system when there is
negligible quantum-mechanical interference between these histories
 as a consequence of the system's initial condition and
dynamics.  Such sets of histories are said to {\it decohere}.

To describe more precisely what is meant by decoherence, consider for
simplicity, a pure initial state $|\psi\rangle$ and a set of histories
defined by sets of alternatives at definite moments of time $t_1,
\cdots, t_n$. The alternatives at the moment of time $t_k$
 are described by an
exhaustive set of exclusive Schr\"odinger  picture projection operators
$\{P^k_{\alpha_k}\}$ satisfying
\begin{equation}
\sum\nolimits_{\alpha_k} P^k_{\alpha_k} = 1, \quad P^k_{\alpha_k}
\ P^k_{\beta_k} =
\delta _{\alpha_k\beta_k} P^k_{\alpha_k}\, . 
\label{fourtwo}
\end{equation}
In this notation $k$ denotes the set of alternatives at time $t_k$
 (\eg a set of
position ranges or a set of momentum ranges, etc.) and $\alpha_k$ the
particular alternative.  The alternatives are {\it fine-grained} if the
$P$'s project onto the one-dimensional subspaces defined by a complete
set of states and otherwise are {\it
coarse-grained}.

The sequences of alternatives at definite moments of time $0<t_1 <t_2 <
\cdots < t_n <T$ define a set of coarse-grained alternative histories on
a time interval $[0,T]$.  The individual histories correspond to
particular sequences $\alpha = (\alpha_1, \cdots, \alpha_n)$ which are
represented by the corresponding chains of projection operators
interrupted by unitary evolution
\begin{equation}
C_\alpha = e^{-iH(T-t_n)/\hbar} P^n_{\alpha_n}
\ e^{-iH(t_n-t_{n-1})/\hbar} P^{n-1}_{\alpha_{n-1}} \cdots
e^{-iH(t_2-t_1)/\hbar} P^1_{\alpha_1} e^{-iHt_1/\hbar}\, .
\label{fourthree}
\end{equation}
These may be written somewhat more compactly using Heisenberg picture
projections as
\begin{equation}
C_\alpha = e^{-iHT/\hbar} P^n_{\alpha_n} (t_n) \cdots P^1_{\alpha_1}
(t_1)\, , \label{fourfour}
\end{equation}
where
\begin{equation}
P^k_{\alpha_k}(t) = e^{iHt/\hbar} P^k_{\alpha_k} e^{-iHt/\hbar}\, .
\label{fourfive}
\end{equation}

Since $\sum_\alpha C_\alpha = e^{iHT/\hbar}$ as a consequence of
\eqref{fourtwo} and \eqref{fourthree}, the evolution of the
initial state $|\psi\rangle$ may be resolved
into branches, $|\psi_\alpha\rangle$,
corresponding to the individual histories
\begin{equation}
e^{-iHT/\hbar}\bigr| \psi \bigr\rangle = \sum\nolimits_\alpha C_\alpha
\bigl| \psi\bigr\rangle \equiv \sum\nolimits_\alpha |\psi_\alpha\rangle
\, . \label{foursix}
\end{equation}
A set of histories decoheres\footnote{The term ``decoherence'' is
used in several different ways in the literature.  We have followed our
earlier work \cite{GH90} in using the term to refer to a property of
a set of coarse-grained {\it histories} of a closed system namely the
absence of interference between individual histories in the set at a
level to ensure the consistency of the probability sum rules.  There are
several different measures of decoherence \cite{GH90b}.  The condition
\eqref{fourseven} is called the ``medium decoherence condition''.  In this
simplified presentation ``decoherence'' therefore means the medium
decoherence of histories.  Similar conditions are called
 ``consistency conditions'' \cite{Gri84, Omn88}
 or ``no-interference conditions'' \cite{YT91b}. The term
decoherence has also been used to refer to the approach to diagonality
of a reduced density matrix in a particular basis.  The decoherence of
density matrices is not the same as the decoherence of histories in
general but the two ideas can be related in special models.}
when the individual branches are essentially
orthogonal
\begin{equation}
\bigl\langle \psi_{\alpha^\prime} \bigl| \psi_\alpha
\bigr\rangle \approx 0\ ,\ \alpha^\prime \not= \alpha\, .
\label{fourseven}
\end{equation}
The probabilities of the individual histories in such a decoherent set are
the square of the norms of the corresponding branches
\begin{equation}
p (\alpha) = \big\Arrowvert
 \bigr| \psi_\alpha
\bigr\rangle\big\Arrowvert^2\, . 
\label{foureight}
\end{equation}

Eq.~\eqref{foureight} is a consistent assignment of probabilities to a
decoherent set of histories because decoherence implies that  the
probability sum rules are satisfied in their most general form.
To give a simple example, consider a set of
histories defined by alternatives at just two moments of time
$t_1$ and $t_2$ and the probability sum rule
\begin{equation}
\sum\nolimits_{\alpha_1} p (\alpha_2, \alpha_1) = p (\alpha_2)\, .
\label{fournine}
\end{equation}
For the left hand side of \eqref{fournine} we may write
\begin{eqnarray}
\sum\nolimits_{\alpha_1} p (\alpha_2, \alpha_1) & = &
\sum\nolimits_{\alpha_1} \bigl\langle \psi \bigl| P^1_{\alpha_1} (t_1)
P^2_{\alpha_2} (t_2) \cdot P^2_{\alpha_2} (t_2) P^1_{\alpha_1} (t_1)
\bigr|\psi \bigr\rangle\nonumber\\
& = & \sum\nolimits_{\alpha^\prime_1\alpha_1} \bigl\langle \psi \bigl|
P^1_{\alpha^\prime_1} (t_1) P^2_{\alpha_2} (t_2) \cdot P^2_{\alpha_2}
(t_2) P^1_{\alpha_1} (t_1) \bigr|\psi\bigr\rangle\nonumber\\
& = & \bigl\langle \psi \bigl| P^2_{\alpha_2} (t_2) \cdot P^2_{\alpha_2}
(t_2) \bigr| \psi \bigr\rangle = p(\alpha_2)\, . 
\label{foureleven}
\end{eqnarray}
The first and last equality are the definition \eqref{foureight}. The second
equality is true because of the decoherence condition \eqref{fourseven}, and
the third because of \eqref{fourtwo}.  Thus, decoherence implies the
probability sum rules needed for a consistent assignment of
probabilities.

Measured alternatives decohere but an alternative need not be a
participant in a measurement situation in order to decohere.  In
cosmology, quantum theory predicts the probabilities of alternative sizes
of density fluctuations one minute after the big bang, in a universe in
which these alternatives decohere, whether or not anything like a
measurement was carried out on them and certainly whether or not there
was an observer around to do it. Decoherence is thus a more precise,
more general, and more observer-independent notion than measurement
and replaces it in the quantum mechanics of closed systems as the criterion
determining which sets of histories can be assigned probabilities.

\section{A Generalized Sum-Over-Histories Quantum\\
Mechanics for Spacetime Alternatives}

Building on the sum-over-histories ideas, the quantum mechanics of
alternatives at definite
moments of time that was described in the previous section may be
generalized to deal with the spacetime alternatives described in
Section III.  The
result is a quantum framework for prediction with dynamics and
alternatives fully in spacetime form.  We shall describe this
generalization for the case of a configuration space spanned by
coordinates, $q^i$, and assume a pure initial state.  There is
nothing essential about these restrictions.  The generalization to an
initial density matrix requires only a modest expansion of the formalism
and the coordinates $q^i$ could be the values of fields
$\phi(\vec x)$ at each point in space.  The important assumption is that
there is a fixed background spacetime that supplies a well defined
notion of time. We follow the discussion in \cite{Har91b}.

The most refined possible description of a closed system are its
fine-grained histories.  These are the paths $q^i(t)$ that are
single-valued functions of the time.  Partitions of these
fine-grained histories into exhaustive sets of exclusive classes
$c_\alpha$ yield sets of coarse-grained histories. An individual
coarse-grained history $c_\alpha$  is thus a class of fine-grained
histories.  Exhaustive sets of alternative coarse-grained histories
$\{c_\alpha\}$ are the general notion of ``observable'' for which
quantum theory predicts probabilities when the set of coarse-grained
histories is decoherent.

\begin{figure}[t]
\begin{center}
\includegraphics[width=4in]{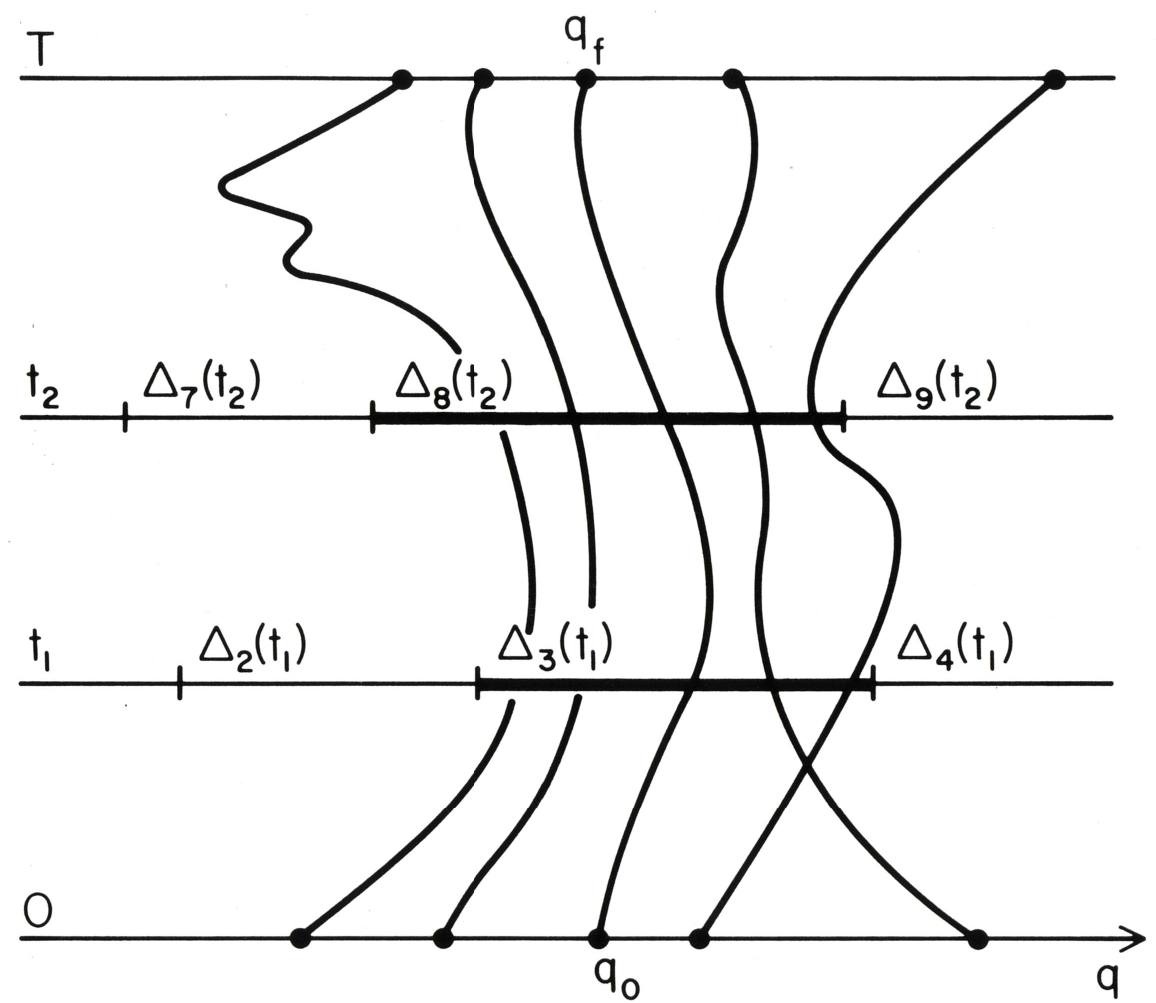}
\caption{Coarse graining by regions of configuration space at
successive moments of time.  The figure shows a spacetime that is a
product of a one-dimensional configuration space $(q)$ and the time
interval $[0,T]$.  At times $t_1$ and $t_2$ the configuration space is
divided into exhaustive sets of non-overlapping intervals:
$\{\Delta_{\alpha_1}(t_1)\}$ at time $t_1$, and
$\{\Delta_{\alpha_2}(t_2)\}$ at time $t_2$ (written in the text as
$\{\Delta^1_{\alpha_1}\}, \{\Delta^2_{\alpha_2}\}$, etc).
  Some of these intervals are
illustrated.  The fine-grained histories are the paths which pass
between $t=0$ and $t=T$.  Because the paths are assumed to be
single-valued in time, the set of fine-grained histories may be
partitioned into two classes according to which intervals they pass
through at times $t_1$ and $t_2$.  The figure illustrates a few
representative paths in the class  which pass through region
$\Delta_3(t_1)$ at time $t_1$ and region $\Delta_8(t_2)$ at time $t_2$.}
\end{center}
\end{figure}

Sequences of alternative ranges of coordinates
$\{\Delta^1_{\alpha_1}\}, \{\Delta^2_{\alpha_2}\}, \cdots,
\{\Delta^n_{\alpha_n}\}$ at, say, times $t_1, \cdots, t_n$
define one kind of partition of the paths $q^i(t)$.  An individual
coarse-grained history corresponding to the sequence of alternatives
 $\alpha = (\alpha_1, \cdots
\alpha_n)$ is the class of paths $c_\alpha$ that thread the region
$\Delta^1_{\alpha_1}$ at time $t_1$, $\Delta^2_{\alpha_2}$ at time
$t_2$, etc. (Figure 3)~~
  These correspond to
sequences of the familiar ``observables'' at definite moments of
time.
However, much more general partitions are possible.  for example,
following the discussion of Section III, paths may be partitioned into
two classes according to
 their behavior with
respect to a spacetime region $R$. One class, $c_0$, consists of all
paths that never intersect $R$ and the other, $c_1$, consists of all
paths that intersect $R$ at least once.  These classes are exclusive and
together they are exhaustive. They  constitute a
coarse-grained set of histories.

The most general notion of coarse-grained histories is a
partition by values of  functionals of histories \cite{Cav86, Harpp}.
Consider, for example, just a
single functional $F[q(\tau)]$ and a set of ranges $\{\Delta_\alpha\}$
of the real line.  The class $c_\alpha$  consists of all paths $q^i(t)$
such that $F[q(\tau)]$ lies in the range $\Delta_\alpha$.  We could
partition the paths, for example, by the value of some particular
coordinate, $q^k$,
averaged over a time interval, that is, by the functional
\begin{equation}
F[q(\tau)] = \frac{1}{T} \int^T_0 dt\ q^k (t)\, .
\label{fiveone}
\end{equation}
In this way we could deal with the average values of fields over
spacetime regions whose importance was stressed by Bohr and
Rosenfeld \cite{BR33}.
In a field theory with a spinor field $\psi(x)$ we could partition the
field histories by the values of currents, \eg $\psi^\dagger (\vec x,t) \psi
(\vec x, t)$.  In this way the theory can incorporate observables
associated with spin.  In the theory of a non-relativistic particle we
could partition the paths by the value of the position {\it difference}
between two times separated by a time interval $T$. In the limit that
$T$ becomes large but still short compared to dynamical time scales of
any interaction these partitions define momentum alternatives determined
by the time of flight \cite{FH65}.
In this way probabilities for momenta can be
predicted by the theory.

Sums over the fine-grained histories contained in a coarse-grained
history define the branch of the initial state corresponding to that
history.  To make this explicit, consider histories on the
time interval $[0,T]$ and
suppose that the fine-grained histories on this interval are partitioned
into an exhaustive set of exclusive classes $c_\alpha$.  We define
branches of the initial state and class operators $C_\alpha$
 corresponding to each class by
\begin{subequations}
\label{fivetwo}
\begin{equation}
\big|\psi_\alpha\big\rangle \equiv C_\alpha\big| \psi \bigr\rangle =
\int\nolimits_{c_\alpha} \delta q
\ \exp \bigl(iS[q(\tau)]/\hbar\bigr) \big|\psi\bigr\rangle
\label{fivetwo a}
\end{equation}
where the integral is over all paths in the class $c_\alpha$. We are
using the same abbreviated notation as in \eqref{twofour}.  More explictly,
\eqref{fivetwo a} stands for
\begin{equation}
\psi_\alpha(q, T) = \bigl\langle q \big|C_\alpha\big| \psi\bigr\rangle
= \int dq^\prime \int\nolimits_{[q^\prime c_\alpha q]} \delta q
\ \exp\Bigl(iS\bigl[q(\tau)\bigr]/\hbar\Bigr) \psi(q^\prime, 0)\, .
\label{fivetwo b}
\end{equation}
\end{subequations}
Evidently, since
the sum over all paths just gives unitary evolution as in \eqref{twofour},
we have
\begin{equation}
\sum\nolimits_\alpha C_\alpha = e^{-iHT/\hbar}\, . 
\label{fivethree}
\end{equation}

Decoherence and probabilities are defined as before. The set of
coarse-grained histories $\{c_\alpha\}$ decoheres if
\begin{equation}
\bigl\langle\psi_{\alpha^\prime}\big|\psi_\alpha\bigr\rangle
\approx 0\quad \alpha^\prime\not=\alpha 
\label{fivefour}
\end{equation}
and the probability of an individual history $c_\alpha$ in
a decoherent coarse-grained set is
\begin{equation}
p(c_\alpha) = \big\Arrowvert  \big| \psi_\alpha \bigr\rangle
\big\Arrowvert^2\, .
\label{fivefive}
\end{equation}

The important point about this construction is that the probability sum
rules, which are the obstacle to consistently assigning probabilities in
quantum mechanics, are satisfied for a decoherent
 set of coarse-grained
histories as a consequence of \eqref{fivefour}.  To see this consider a
partition of the $\{c_\alpha\}$ into an exhaustive set of exclusive
classes yielding a coarser-grained partition of the
fine-grained histories  $\{\bar c_\beta\}$.
The most general form of the probability sum
rules is
\begin{equation}
p(\bar c_\beta) = \sum_{\alpha\epsilon\beta} p (c_\alpha)
\label{fivesix}
\end{equation}
where the sum is over $\alpha$ such that $c_\alpha$ included in $\bar
c_\beta$.  This is easily seen to be satisfied as a consequence of
\eqref{fivetwo} and \eqref{fivefour}.  From the linearity of
\eqref{fivetwo} that
reflects the principle of superposition it follows that
\begin{equation}
\bar C_\beta = \sum_{\alpha\epsilon\beta} C_\alpha\, .
\label{fiveseven}
\end{equation}
But then a repetition of the argument that led to \eqref{fournine} shows the
validity of the more general sum rule \eqref{fivesix}.
In this way sum-over-histories ideas can be used to formulate a
generalized quantum mechanics of closed systems that is fully in
spacetime form with dynamics summarized
by path integrals over alternatives that are general partitions of
spacetime histories.

\section{Comparison with Schr\"odinger-Heisenberg\\
Quantum Mechanics}

We mentioned that, for alternatives at moments of time,
sum-over-histories quantum mechanics deals directly
only with configuration space
alternatives in contrast to Schr\"odinger-Heisenberg quantum mechanics
which utilizes all the possibilities of transformation theory.
The sum-over-histories formulation is thus more
restricted in its alternatives at moments of time.
Configuration space variables have a
preferred place in the formalism.

However, in the preceeding section we have shown how sum-over-histories
ideas can be used to formulate a quantum mechanics that deals with
spacetime alternatives that are much more general than those ``at
moments of time'' with which the Schr\"odinger-Heisenberg formulation is
concerned.  This generalized  spacetime quantum mechanics cannot be
reformulated in terms of the two laws of evolution of the
Schr\"odinger-Heisenberg formulation.  If it could be so formulated, the
class operators $C_\alpha$ defined by \eqref{fivetwo} for each history in a
spacetime coarse graining could be represented as a chain of projections
of the form \eqref{fourfour} or perhaps a continuous product of such
projections one for each time.  Then one could describe the
evolution completely in terms of unitary evolution interrupted (perhaps
continuously in time) by reductions of the state vectors.

However,
the class operators of a spacetime coarse graining cannot, in
general, be represented as products of projections, even continuous
ones.  Consider, by way of example, the coarse
graining of the paths of a particle by their behavior with respect to a
spacetime region $R$ that was discussed earlier.  The class operator
$C_0$ for the class of paths that never cross $R$ {\it can} be
represented as a continuous chain of projections on the region of $q$
outside of $R$ at each time.  However, the class operator $C_1$ for the
class of paths that crosses $R$ sometime cannot be so represented
because at each time the particle could be either
inside $R$  or outside it.
The class operator $C_1$ can be represented formally as a {\it sum} of
continuous chain of projections
\begin{equation}
C_\alpha = \sum_{\alpha(t)\epsilon c_\alpha} \prod_t
\, P^{k(t)}_{\alpha(t)} (t)\, .
\label{sixone}
\end{equation}
However, a quantum mechanics of alternatives represented by
class operators that are {\it sums} of chains of projections
already constitutes
a generalization of familiar quantum mechanics \cite{Har91,
GH92} although a very natural one.

The probabilities of spacetime coarse-grained histories thus cannot be
expressed in terms of a unitarily evolving state vector that is
``reduced'' at various moments of time.  The reason, it should be
stressed, does not lie in the
use of path integrals versus operators.  Indeed, as
\eqref{fivetwo} shows there is a correspondence between path integrals and
operators and
operator techniques provide the most convenient way of defining the path
integrals on the right hand side of \eqref{fivetwo} \cite{Har91b}.
Rather, it is the
space{\it time} nature of the alternatives that does not allow
meaningful notion of state of the system at a moment of time.

This generalized sum-over-histories quantum mechanics is thus not
equivalent to the Schr\"odinger-Heisenberg formulation because the two
formulations deal with different alternatives.  Sums of continuous
products like \eqref{sixone} could be taken as the starting point of a yet
more generalized quantum mechanics that would contain all the
alternatives of both.  Such a generalization presents interesting
mathematical problems.  In the meantime,
to the extent our
experience can be expressed in terms of spacetime alternatives, nothing
seems to be lost by a restriction to these and much would seem to be
gained from a more realistic description of alternatives are
extended over time.  In the next section I shall argue that the spacetime
approach to quantum mechanics has definite advantages in the quantum
theory of spacetime where there is no well defined notion of ``at a
moment of time''.

\section{A Generalized Quantum Mechanics of Spacetime}

The nature of the ``observables'' to which a quantum theory of
spacetime geometry assigns probabilities has always been
something of a puzzle in quantum gravity.  We cannot straightforwardly
and covariantly define alternatives at a moment of time
because there is no fixed notion of time.  In a theory where the
geometry of spacetime fluctuates quantum-mechanically, there is no fixed
interval between spacetime points and not even a fixed notion of whether
that interval is timelike or spacelike.  Put differently, there is no
covariant choice of geometrical varible to play the role of ``$t$'' in
the predictive framework summarized by \eqref{fourfour}, \eqref{fourseven},
 and \eqref{foureight}. In
essence, there is a conflict between the diffeomorphism
invariance of spacetime theories of gravity and the requirements of
Schr\"odinger-Heisenberg quantum mechanics.  This is called
the ``problem of time'' in quantum gravity.
For recent critical surveys of various approaches to its resolution
see \cite{Kuc81,Ishpp, Kuc92, Unrpp}.

One thread of thought is that alternatives in quantum gravity should be
represented by operators that commute with the constraints implied by
diffeomorphism invariance.  However, time reparametizations are among
the diffeomorphisms and for such theories the
constraints generate the dynamics.  Restricting the observables to
operators that commute with the constraints corresponds classically to
observables that are constants of the motion.
This is a
very restricted class of observables!\footnote{However, argued
to be sufficient by some, Rovelli \cite{Rov90}.}

The spacetime approach to quantum mechanics provides a different
resolution to the problem of time.  In a quantum theory of spacetime, the
fine-grained histories are the possible
four-dimensional metrics and matter field
configurations on a fixed manifold $M$ (in the simplest case).  Quantum
gravitational dynamics can be expressed in spacetime form through sums
of $\exp(iS[g,\phi]/\hbar)$ over metrics and matter fields on $M$, where
$S$ is the action for spacetime and matter.  The use of
sums-over-histories to formulate a covariant quantum gravitational
dynamics has been extensively investigated.\footnote{It has been
investigated in a literature far too large to be cited here.  However,
some representative and important early papers are those of Misner
\cite{Mis57}, Leutwyler \cite{Leu64},
DeWitt \cite{DeW79}, Fradkin and Vilkovisky \cite{FV73},
Faddeev and Popov \cite{FP74},
Hawking \cite{Haw79}, Teitelboim \cite{Tei83}, and Polyakov
\cite{Pol81}.  There are many other important ones.}
However, the spacetime approach to quantum mechanics can also be used to
specify the alternatives to which a quantum theory of spacetime assigns
probabilities in a covariant way.

A generalized sum-over-histories quantum mechanics for spacetime can be
constructed in which the fine-grained histories are the four-dimensional
metrics and matter field configurations as discussed
above \cite{Har91, Harpp}. Allowed coarse grainings are
partitions of these fine-grained histories into exhaustive sets of {\it
diffeomorphism} {\it invariant} classes $\{c_\alpha\}$.  These
diffeomorphism invariant classes are the analogs of the spacetime coarse
grainings we have discussed for quantum mechanics in fixed background
geometries. They supply a very broad class of generally covariant
alternatives to which, when decoherent, a quantum theory of spacetime
will assign probabilities.

Many examples of diffeomorphism invariant coarse grainings could be
given but in the limited space available I will confine myself to just
two.  First, we consider the probability that a closed cosmology reaches
a maximum spatial volume greater than, say, $V_0$.  This question can be
given a precise meaning by partitioning the class of all cosmological
four-metrics into the class that have at least one spacelike
three-surface with a total volume greater than $V_0$, and the class of
three metrics that have no such three-surface.  This is clearly an
exhaustive partition into two exclusive classes that are each
diffeomorphism invariant.  Equally clearly the specification of these
alternatives involves no preferred notion time.

\begin{figure}[t]
\begin{center}
\includegraphics[width=5in]{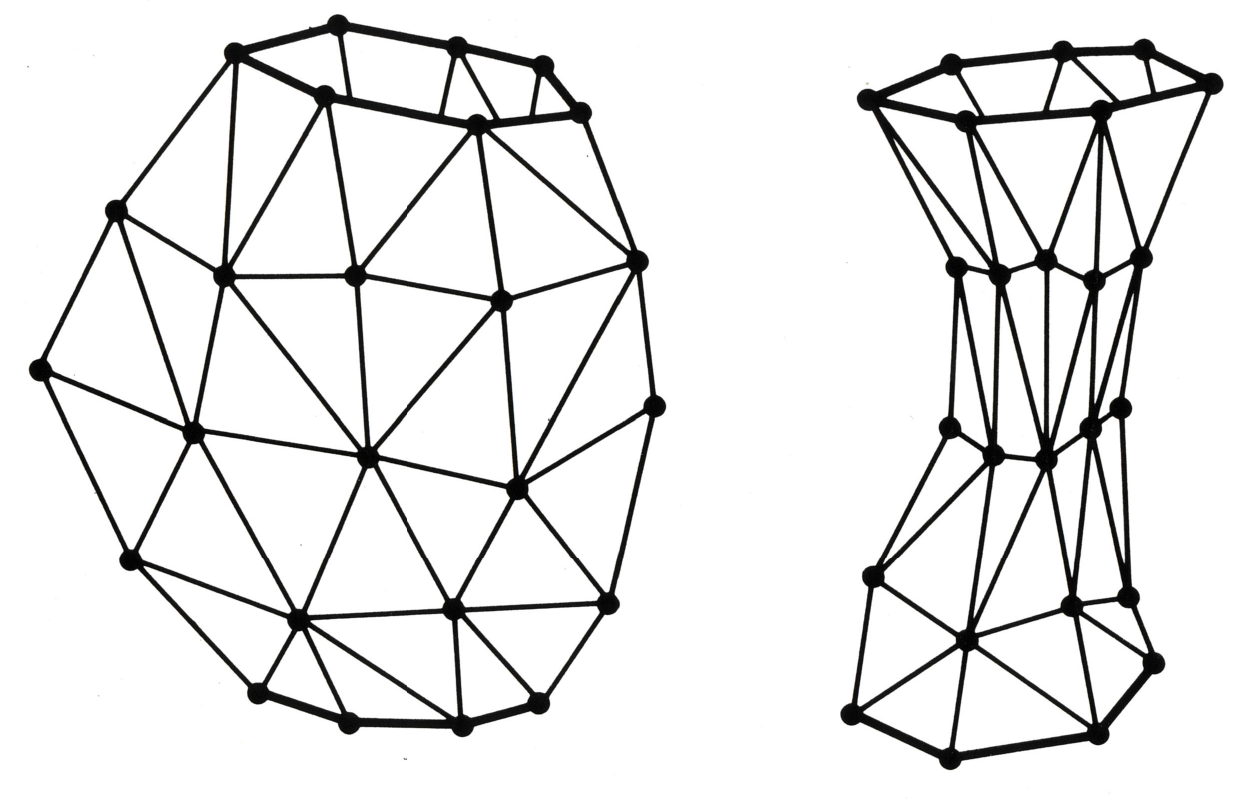}
\caption{Simplicial Geometries.
Two-dimensional surfaces can be made up by joining together flat
trangles to form a simplicial namifold.  A geometry of the surface is
specified by an assignment of squared edge-lengths to the triangles.  The
figure shows two different geometries obtained by a different assignment
of squared edge-lengths to the same simplicial manifold.  The
generalization of these ideas to four dimensions and Lorentz signature
gives the natural lattice version of general relativity --- the Regge
calculus.  In a sum-over-histories quantum theory of simplicial
spacetimes, sums over geometries are represented by integrals over the
squared edge-lengths.  Diffeomorphism invariant alternatives can be
defined by partitioning the space of allowed squared edge-lengths into
exhaustive sets of exclusive regions.  For example, one could partition
closed cosmological geometries into the class that has no simplicial
spacelike three surface greater than a certain volume and the class that
has at least one such surface.  In a given simplicial manifold it is
possible to enumerate all three surfaces and identify the regions in the
space of squared edge-lengths to which each class corresponds.}
\end{center}
\end{figure}

As a second example, we consider the natural lattice formulation of
general relativity --- the Regge calculus \cite{Reg61, Wil92}.  A
two-dimensional surface can be constructed from triangles (Figure 4).
The topology of the surface is specified by how the triangles are joined
together.  The geometry of the surface is specified by an assignment of
the squared edge-lengths of the triangles and a flat geometry to their
interior.  Similarly, four-dimensional Lorentzian geometries can be
constructed out of four-simplices with squared edge-lengths that may be
positive or negative.  The space of geometries is parametrized by the
$n_1$ squared edge-lengths $s^1, \cdots, s^{n_1}$
consistent with the analogs of triangle inequalities.  A point in this
space of squared edge-lengths
is a fine-grained history.  A partition of this space
   into regions that are invariant under the symmetries of
the simplicial net provides a very general class of coarse grainings for
these lattice geometries that does not require a preferred notion of
time for its specification.

Beyond these examples, however, diffeomorphism invariant partitions of
metrics and field configurations supply the {\it most general} notion of
alternative in quantum theory that is describable in spacetime terms.
Every property of the universe whose probability we may seek to
calculate corresponds to some such coarse-graining, namely the partition
of all four-metrics and matter field configurations into the class in
which the property is true and the class in which it is false.  If we
cannot distinguish which fine-grained histories have the property and
which do not then the property is not well defined.

To complete the construction of a generalized quantum mechanics a notion
of decoherence must be specified for these coarse grained histories.  I
shall only indicate how to do this schematically.  More details can be
found in \cite{Harpp}.  Branches, corresponding to individual
coarse-grained histories in a set $\{c_\alpha\}$, can be represented as
wave functions $\Psi_\alpha[h,\chi]$ on the superspace of spatial three
metrics, $h_{ij}({\bf x})$, and spatial matter field configurations,
$\chi{(\bf x)}$.  These are defined in analogy to \eqref{fivetwo} or,
more explicitly, \eqref{twofive}
\begin{equation}
\Psi_\alpha [h, \chi] = \int\delta h^\prime\delta \chi^\prime
\int\nolimits_{[(h^\prime, \chi^\prime), c_\alpha,
(h, \chi)]} \delta g\,\delta\phi\,\exp
\Bigl(iS\bigl[g,\phi\bigr]/\hbar\Bigr)\circ
\Psi\left[h^\prime,\chi^\prime\right]\, .
\label{sevenone}
\end{equation}
In this expression $\Psi[h,\chi]$ is the wave function representing the
initial condition of the universe, say, the ``no-boundary'' wave
function \cite{HH83}.  The integral is over metrics and fields in the class
$c_\alpha$ on a manifold $M$ with two boundaries.  On one boundary
metrics and fields match the arguments $(h,\chi)$ of the branch wave
function.  On the other boundary they match the arguments $(h^\prime,
\chi^\prime)$ of the initial condition.  Much remains to be spelled out
to make such a construction concrete, not least the details of the
measure and the product $\circ$ with which the initial condition is
attached to the functional integral and  more details are in
\cite{Harpp}.  The important point is that with these branches one can
define a decoherence condition for coarse-grained histories analogous to
\eqref{fivefour} and probabilities for the individual histories in decoherent
sets by expressions analogous to \eqref{fivefive}. The quantum mechanics of
spacetime is thus cast into a generaly covariant fully spacetime form
free from the problem of time.

How is the usual formulation of quantum mechanics with its preferred
time variables connected with this generalization that requires no such
preferred time? The answer is to be found by examining the origin of the
classical spacetime of everyday experience.  Classical spacetime is not
a general feature of every state in a quantum theory of gravity.  We
expect classically behaving spacetime only for particular states and
then only for coarse-grainings that define geometry well above the
Planck scale.  For such states and coarse-grainings the integral over
metrics in
\eqref{sevenone} may be carried out by steepest descents.  Suppose, in the
simplest case, that only a single classical geometry $\hat g$ dominates
the sum.  The remaining integrals over matter fields are equivalent to
those of a field theory in the fixed background geometry of $\hat g$.
The quantum mechanics of matter fields thus inherits its notions of time
from the timelike directions of the classical background $\hat g$. It could
be in this way that the familiar Hamiltonian formulation of quantum
mechanics, with its preferred time(s),
 emerges as an approximation appropriate to the existence of an
approximately quasiclassical spacetime in a more general, covariant,
spacetime, formulation of quantum theory that is
free from the problem of time.

\acknowledgments

The author has benefited from many conversations with many physicists on
these subjects, but especially with M.~Gell-Mann, on the quantum
mechanics of closed systems, with K.~Kucha\v r on the problem of time in
quantum gravity, and with R.~Sorkin on sum-over-histories formulations of
quantum mechanics.  Preparation of this report as well as the work it
describes was supported, in part, by the National Science Foundation
under grant PHY90-08502.

\end{document}